\documentclass[11pt,a4paper]{article}
\pdfoutput=1

\usepackage{jcappub}
\usepackage{amsmath}
\usepackage{amsthm}
\usepackage{graphicx}
\usepackage{amsfonts}

\title{Quantum Hubble horizon}

\author{Micha{\l}~Artymowski${}^{ab}$} 
\author{and Jakub~Mielczarek${}^{bc}$}

\affiliation{$^a$Institute of Theoretical Physics, Faculty of Physics, University of Warsaw ul. Pasteura 5,
02-093 Warsaw, Poland}
\affiliation{$^b$Institute of Physics, Jagiellonian University, {\L}ojasiewicza 11, 30-348 Cracow, Poland}
\affiliation{$^c$Aix Marseille Univ, Universit\'e de Toulon, CNRS, CPT, Marseille, France}

\emailAdd{michal.artymowski@uj.edu.pl}
\emailAdd{jakub.mielczarek@uj.edu.pl}

\newcommand{\red}{\textcolor{\red}}

\abstract{
The article addresses a possibility of obtaining cosmologically relevant effects from the 
quantum nature of the Hubble horizon. Following the observation made by E.~Bianchi and 
C.~Rovelli in Phys.\ Rev.\ D {\bf 84} (2011) 027502 we explore relationship between the 
Planck scale discreteness of the Hubble horizon and deformations of the symmetry of 
rotations. We show that the so-called $q$-deformations in a natural way lead to a mechanism 
of condensation in the very early Universe. We argue that this provides a possible resolution 
of the problem of initial homogeneity at the onset of inflation.  Furthermore, we perform 
entropic analysis of the quantum Hubble horizon and show that the $\Lambda$CDM model 
may arise from linearly (in area of the horizon) corrected Bekenstein-Hawking entropy. 
Based on this, we have shown that the current accelerating expansion can be associated 
with the entropy decrease in the Hubble volume. The presented results open new ways to 
explore relation between the Planck scale effects and observationally relevant features of 
our Universe. 
}

\begin{document}

\maketitle

\section{Introduction}

Cosmic inflation \cite{Guth:1980zm,Starobinsky:1980te,Lyth:1998xn,Liddle:2000dt} is a hypothetical 
period of the evolution of the early Universe characterized by the accelerated expansion of space. 
It is a powerful theoretical tool to solve problems of the classical Big Bang cosmology, such as horizon 
or curvature problem. Recent observational data from the Planck satellite \cite{Ade:2013zuv} has 
set new upper bounds on the tensor-to-scalar ratio $r$, which in inflationary theories is proportional 
to the slow-roll parameter $\epsilon$. The upper bound for $r$ sets the GUT scale to be the maximal 
scale of inflation at the moment of the horizon crossing of the pivot scale, i.e. around 60 $e$-folds 
before the end of inflation. Nevertheless, inflation could in principle start in much higher scales, up 
to the Planck scale. 

The idea of a high scale of the beginning of inflation is well motivated theoretically within the approach 
of quantum tunneling of the Universe around the Planck scale. Such a Universe is very likely to immediately 
recollapse unless the Planckian Universe is in the quantum state that mimics the cosmological constant 
with the equation of state $p \simeq -\rho$ \cite{Brout:1979bd,Grishchuk:1981xf,Vilenkin:1982de,Mukhanov:2014uwa}. 
The high scale of the beginning of inflation has recently also been supported by the results from 
the theory of Causal Dynamical Triangulations \cite{Ambjorn:2012jv}. The cosmological constant 
(which in the realistic case should be replaced by the inflationary potential) plays the crucial role 
in the process of creating a classical Universe from the ``quantum foam''. 

The other argument to start inflation close to the quantum gravity scale is the problem of initial 
conditions mentioned in Ref. \cite{Ijjas:2013vea}, which points out that inflationary models like 
Higgs \cite{Bezrukov:2007ep} or Starobinsky inflation \cite{Starobinsky:1980te} have massively 
finely tuned initial conditions \footnote{Note that the problem of initial conditions is still discussed 
within the scientific community. For opposite points of view see e.g. Refs \cite{Linde:2014nna,
Gorbunov:2014ewa,East:2015ggf,Clough:2016ymm,Kleban:2016sqm}. For some of the proposed 
solutions see Refs. \cite{Guth:2013sya,Dalianis:2015fpa,Linde:2004nz,Hamada:2014wna,Carrasco:2015rva,
Artymowski:2016ikw,Dimopoulos:2016yep,Artymowski:2014gea}}. In those models the inflationary part 
of potentials is limited from above by the scale of inflation, which is typically of order of the GUT scale. 
The problem is following - Let us consider a horizon in the pre-inflationary Universe filled with 
inflaton $\phi$, dust and radiation. We assume that at such a high energies the contribution of 
the cosmological constant is negligible. For the plateau potential only the kinetic term $\dot{\phi}^2$ 
and gradient of the field $(\partial^i\phi)^2$ have significant contribution to the inflaton's energy 
density. The energy density of the kinetic term, radiation, dust and gradient decrease like 
$a^{-6}$, $a^{-4}$, $a^{-3}$ and $a^{-2}$ respectively, where $a$ is a scale factor. Therefore, 
the inhomogeneous part of the energy density shall dominate the system before the potential 
term of the inflaton has a chance to generate acceleration of the scale factor, which would strongly 
suppress inhomogeneities. As the result the successful inflation may not happen, unless one 
assumes around $10^9$ homogeneous, causally independent horizons at the Planck scale 
\cite{Ijjas:2013vea}. 

In order to start inflation one usually assumes the existence of the patch of space, which is 
homogeneous enough to support initial conditions for inflation. This part of the Universe would 
be most likely filled with a condensate of a scalar field (or fields), since beside few exceptions the 
cosmic inflation is usually run by the homogeneous scalar field. The creation of a homogeneous 
scalar field is natural on the onset of inflation, when all of the inhomogeneities are exponentially 
suppressed. Nevertheless, the existence of a pre-inflationary Planckian horizon filled with a 
homogeneous scalar field seems to be fine-tuned. In this article, we investigate the possibility of 
a naturalness of the homogeneity of the Planckian Universe in the framework of quantum gravity. 

We show that Planck scale discreteness of the cosmological (Hubble) horizon introduces a possible 
mechanism leading to homogeneous initial conditions at the onset of inflation. The mechanism 
relies on the properties of quantum gravitational effect leading to the noncommutative behavior 
characterized by the so-called $q$-deformations. The vale of $q$-deformation parameters is 
a function of energy density scale and in the very early Universe only limited number of representations 
of the $q$-deformed group is allowed. In consequence, the Universe establishes a condensate 
state when energy densities approach the Planck energy scale. The mechanism is introduced 
Sec. \ref{Condensation}, where we also stress that the $q$-deformations in combination with 
Copernican Principle lead to homogeneity.  Then, in Sec. \ref{Inflation} meaning of the 
performed considerations in the context of the problem of initial homogeneity at the beginning 
of inflation is explained. Furthermore, preliminary analysis of the energy density fluctuations 
of the Hubble horizon are performed. In Sec. \ref{Thermodynamics}, entropic properties of 
the quantum Hubble horizon are analyzed. We show that, depending on the mater content of
the Universe the entropy flow may occur either into or outside of the Hubble volume. The form of 
the entropy as a function the area of the Hubble sphere is reconstructed for the $\Lambda$CDM 
Universe.  The results are summarized and discussed in Sec. \ref{Summary}. Furthermore, in 
the Appendix the issue of $q$-deformations in Loop Quantum Gravity approach to the Planck scale 
physics is outlined.

Throughout this article we consequently apply the Planck units, where $\hslash = c = k_{B}= 1$ and
$G = l_{\text{Pl}}^2$, where $l_{\text{Pl}} $ denotes the Planck length.

\section{Condensation via $q$-deformations} \label{Condensation}

\begin{figure}[ht!]
\centering
\includegraphics[width=6cm,angle=0]{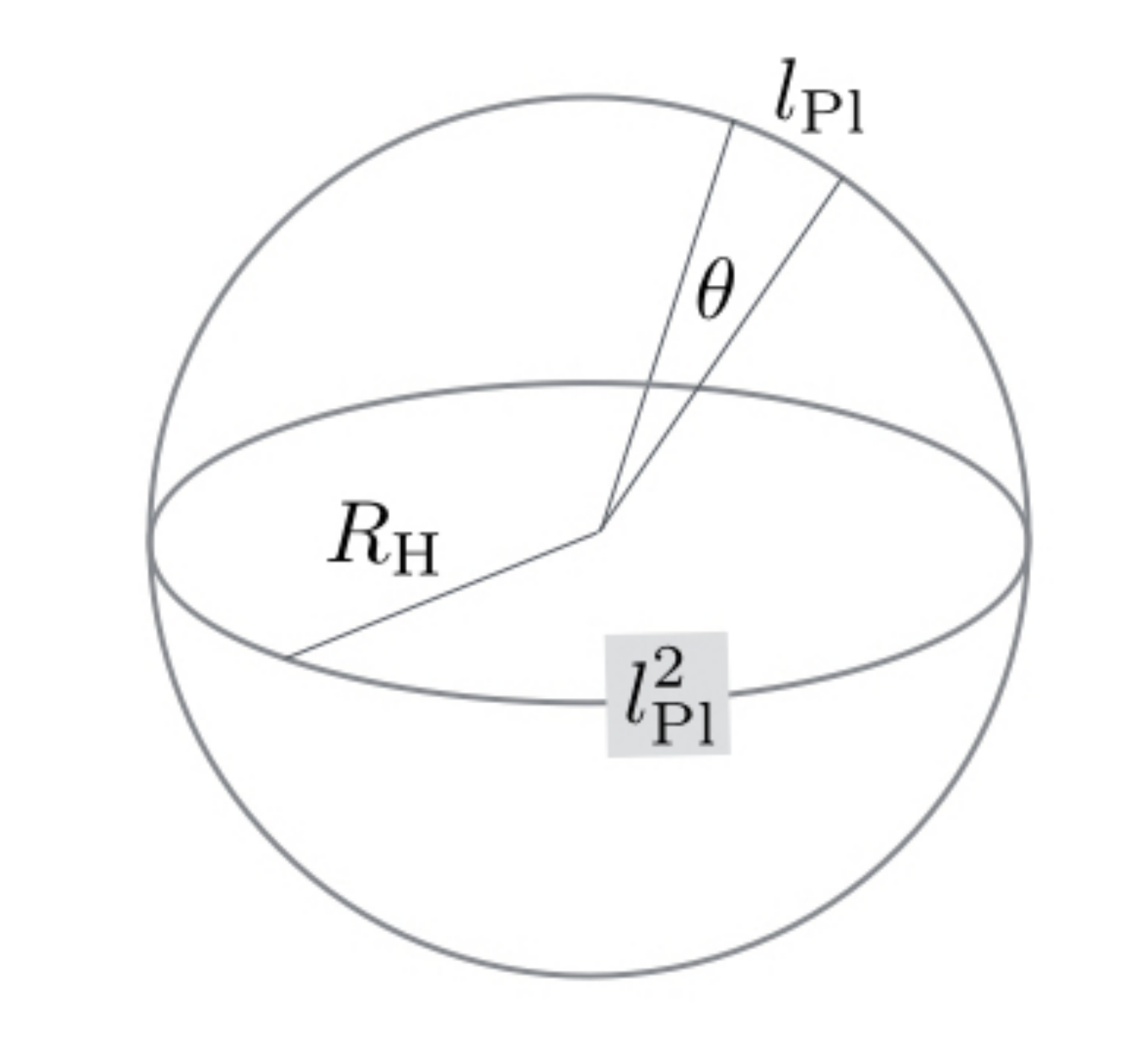}
\caption{Quantum Hubble horizon can be considered as a set of Planckian pixels (a single such a pixel 
is depicted as the shadowed square).
Radius of the sphere is $R_{\text{H}}$ and the maximal angular resolution is denoted as $\theta$.} 
\label{hs}
\end{figure}

One of the characteristic expectations regarding the Planck scale physics is that there is a minimal 
length scale, being of the order of the Planck length $l_{\text{Pl}} \approx 1.62 \cdot 10^{-35}$m. Depending
on the particular model of the Planck scale physics, the Planck length may enter in a various way 
into the considerations. However, the general qualitative conclusion is common -- no details of the 
space-time structure at the scales below the Planck length can be observed. The Planck length, 
therefore, sets the highest (UV) limit on resolution at which space-time can be probed. But, there is 
also the lowest (IR) limit which results from the causal structure of space-time and is given by the 
so-called Hubble radius: 
\begin{equation}
R_{\text{H}} := \frac{1}{H} \label{HubbleRadius}\, , 
\end{equation}
where $H$ is the Hubble factor. The Hubble radius allows to define a Hubble sphere (see Fig. \ref{hs}) containing 
all information accessible to the observer located at the center of the sphere. The sphere has the area 
$A_{\text{H}} =4\pi R^2_{\text{H}}$. While the minimal area allowed by Planckian physics is of the order 
of  $l_{\text{Pl}}^2$, the Hubble sphere contains approximately $A_{\text{H}}/l_{\text{Pl}}^2=
4\pi \left( R_{\text{H}}/l_{\text{Pl}}\right)^2$ elementary Planckian cells (pixels). In the present Universe, 
the number is extremely large:
\begin{equation}
\frac{A_{\text{H}}}{l_{\text{Pl}}^2}=4\pi \left(\frac{R_{\text{H}}}{l_{\text{Pl}}}\right)^2 
\approx 8 \cdot 10^{122},   \label{currentjmax}
\end{equation} 
where we used $R_{\text{H}}= \frac{1}{H_0}  \approx 4,4$ Gpc. Worth mentioning here is that the 
number is also proportional to the number of degrees of freedom stored at the Hubble sphere. 
As we will discuss later, what will be crucial for the mechanism we are going to introduce, the quantity 
analyzed in Eq. (\ref{currentjmax}) dramatically decreases when the Planck epoch of the evolution 
of the Universe if approached, where $R_{\text{H}}$ falls to the value being of the order of $l_{\text{Pl}}$. 

As discussed in Ref. \cite{Bianchi:2011uq}, the Hubble sphere decomposition on the Planckian cells 
leads to the maximal allowed angular resolution given by 
\begin{equation}
\theta \sim \frac{l_{\text{Pl}}}{R_{\text{H}}} \approx 10^{-61}\ \text{rad}, \label{minangle}
\end{equation}
where the numerical value has been given for the current value of $R_{\text{H}}$. Because of 
this, the rotational invariance is not fully satisfied but instead there is a minimal angle given by 
Eq. (\ref{minangle}) below which rotations cannot be considered. In consequence, the rotation 
group $SO(3)$ or its double covering counterpart (the $SU(2)$ group) require an adequate 
modification, taking into account the maximal resolution given by Eq. (\ref{minangle}).

Such modifications are known in mathematical physics under the name of $q$-deformations, 
which in case of rotations lead to the $SU(2)_q$ group, where in general $q \in \mathbb{C}$
\footnote{The case of $q \in \mathbb{R}$ has been introduced in \cite{Woronowicz:1987}.}. In such a case 
the $q$-deformation factor can be written as $q=e^{i\frac{\pi}{k}}$ with $k \in \left\{2, 3, 4, \dots \right\}$, 
which modifies properties of the $SU(2)$ group. In particular, only the following values $j$ (labeling irreducible 
representations of the $SU(2)_q$) are allowed: $j \in \left\{0, \frac{1}{2},1, \dots, \frac{k}{2}-1\right\}$,  
while the ${\rm SU}(2)$ group allows for arbitrary representations labeled by half-integers 
$j = \frac{n}{2} \in \mathbb{N}$. Furthermore, dimensionality of irreducible representations of the $SU(2)_q$ group 
are given by  the following formula:
\begin{equation}
d_j^q = \frac{q^{2j+1}-q^{-(2j+1)} }{q-q^{-1}} =
\frac{\sin \left( \frac{\pi}{k}(2j+1)\right)}{\sin \left( \frac{\pi}{k}\right)}, \label{SU2qdim}
\end{equation} 
such that for $q\rightarrow 1$ ($k \rightarrow \infty$) the  $SU(2)$ case with $d_j=2j+1$ is recovered. 

One can ask: what is the maximal angular resolution corresponding to the case of $q$-deformation 
with some $j_{\text{max}} := \frac{k}{2}-1$? In case of the standard $SU(2)$ group the degeneracy 
of the $j$ representation is equal to $d_j=2j+1$, which means that the corresponding resolution square 
is $\theta^2 \approx \frac{4\pi}{2j+1}$. However, in the $q$-deformed case the dimensionality of the 
representation is not a monotonic function of $j$ and the maximal degeneration corresponds to the 
maximum of the function (\ref{SU2qdim}), which is located at $j_0= \frac{1}{2} j_{\text{max}}$ 
for which $d_{j_0}^q=1/\sin \left( \frac{\pi}{k}\right) \approx \frac{2}{\pi} (j_{\text{max}}+1)$~\footnote{
In Ref. \cite{Bianchi:2011uq} it has been argued that $\theta^2 \approx 2/j_{\text{max}}$}. 
The approximation is valid for sufficiently large values of $j_{\text{max}}$. Anyway, similarly to what is 
expected based on the formula $\theta^2 \approx \frac{4\pi}{2j+1}$,  the resolution square is 
$\theta^2 \sim \frac{1}{j_{\text{max}}+1}$. In consequence, the $\pi/k$ factor entering the expression 
$q=e^{i\frac{\pi}{k}}$ can be written as
\begin{equation}
\frac{\pi}{k} = \frac{\pi}{2(j_{\text{max}}+1)} \sim \theta^2 \sim  \frac{l_{\text{Pl}}^2}{A_{\text{H}}},
\end{equation}
where we employed the formula for the maximal angular resolution (\ref{minangle}). Furthermore, 
(from the definition) area of the Hubble sphere is proportional to the inverse square of the 
Hubble factor and based on the Friedmann equation $H^2 = \frac{8 \pi l^2_{\text{Pl}}}{3} \rho$ we 
can write that $\frac{\pi}{k} \sim \rho/\rho_{\text{Pl}}$, where the Planck energy density 
$\rho_{\text{Pl}} := l_{\text{Pl}}^{-4}$. Based on this, formula for the $q$-deformation 
parameter can be written as:
\begin{equation}
q = \exp \left(i \frac{\pi}{2} \frac{\rho}{\rho_{*}}\right), \label{qdefformulagen}
\end{equation}
where $\rho$ is the total energy density of the Universe (including cosmological constant)
and $\rho_* \sim \rho_{\text{Pl}} $ is an energy scale comparable with the Planck energy density. 
The formula (\ref{qdefformulagen}) is defined such that $\rho_{*} $ is the maximal energy 
density at which $q=i$ and consequently $k=2$ and $j_{\text{max}}=0$. Angular maximal 
resolution square tends to $4\pi$ in this limit, which corresponds to full isotropy.  

Based on the above arguments one can now conclude that because of the Planck scale 
discreteness the rotational symmetry is affected and the magnitude of this effects increases 
together with increase of the energy density in the Universe. While the effect is marginal today, 
when energy density reaches the Planck values the angular resolution decreased dramatically 
which is associated with the reduction of the allowed representations of the group of rotations. 
In the quantum case, the dimensionality of the Hilbert space associated with the $SU(2)_q$
invariant system is decreasing with the increase of the energy density. In the limiting case 
of $\rho=\rho_*$, only the $j=0$ state $ | 0 \rangle $ is allowed.     

Therefore, in case the gravitational or matter degrees of freedom are associated with the angular 
momentum (spin) then such system undergoes condensation as a result of the $q$-deformation,
which prevents excited states to be occupied. In the $q\rightarrow i$ limit the quantum state of 
multiple degrees of freedom quantum system reduces into the ground state: 
\begin{equation}
| 0 \rangle \otimes | 0 \rangle \otimes | 0 \rangle \otimes | 0 \rangle \otimes | 0 \rangle \otimes | 0 \rangle \otimes \dots. 
\end{equation}
The decrees of angular resolution naturally indicates that the configuration is becoming isotropic. 
Furthermore, taking into account the Copernican Cosmological Principle (no point in space is 
preferred), the isotropy implies homogeneity:
\begin{equation}
\text{Isotropy} + \text{Copernican Cosmological Principle} \Rightarrow \text{Homogeneity}.
\end{equation}  

\begin{figure}[ht!]
\centering
\includegraphics[width=10cm,angle=0]{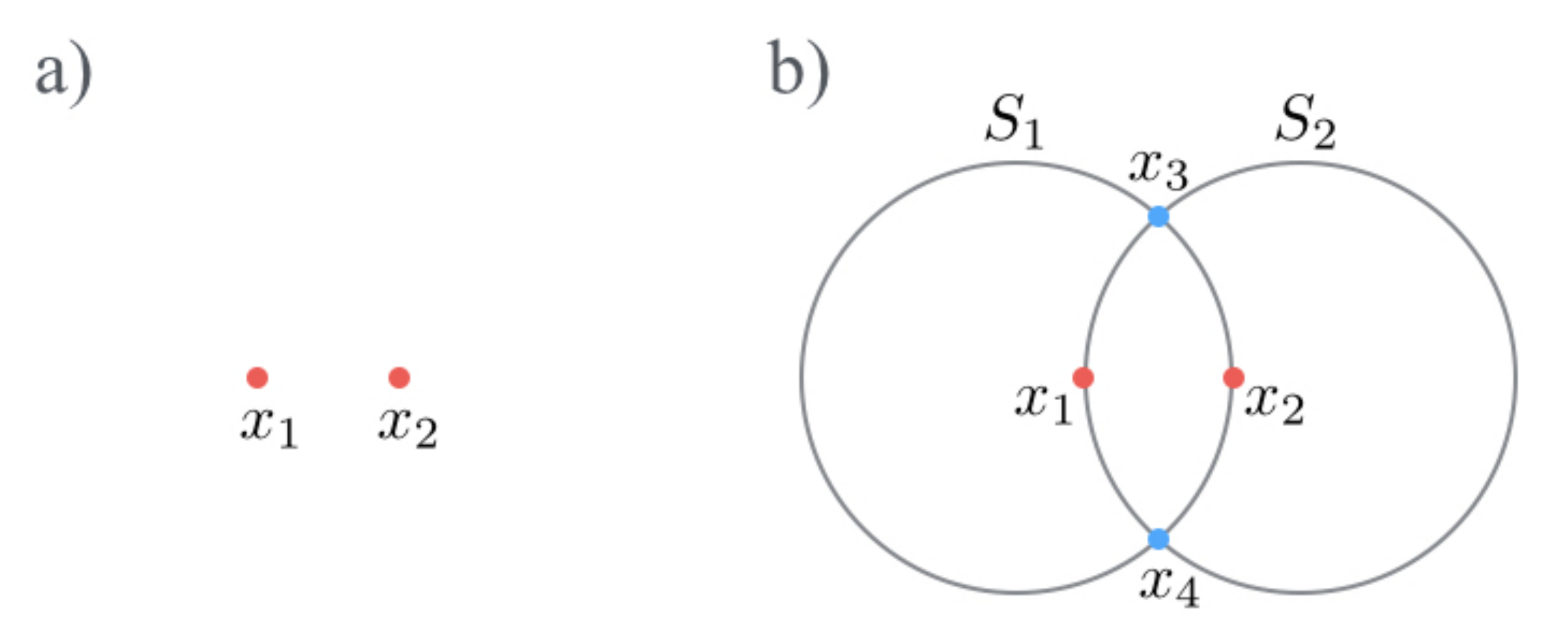}
\caption{a) Arbitrary chosen two points $x_1$ and $x_2$ separated by $d(x_1,x_2) < 2 R_{\text{H}}$. 
b) Geometric constructions used to prove that  $\langle \phi(x_1)\rangle = \langle \phi(x_2)\rangle$ 
assuming statistical isotropy and  Copernican Cosmological Principle.} 
\label{ccp}
\end{figure}

For completeness of our considerations let us give a simple proof of the above statement. For this 
purpose let us consider two points two points $x_1$ and $x_2$ separated by $d(x_1,x_2) < 2 R_{\text{H}}$
(see Fig. \ref{ccp}) and some field $\phi(x)$ which is probed. The task is to prove that for any 
such two point, the statistical isotropy and Copernican Cosmological Principle imply that 
$\langle \phi(x_1)\rangle = \langle \phi(x_2)\rangle$. The averaging is performed either over 
an ensemble of the configurations of the field $\phi(x)$ or over different points.  

The proof can be performed with the use of following geometrical construction: 
Let us consider two Hubble spheres $S_1$ and $S_2$ as shown in Fig. \ref{ccp}.
The statistical istotropy implies that $\langle \phi(x_3)\rangle = \langle \phi(x_4)\rangle$ 
for both observers located at the centers of the Hubble spheres. Now, because $\forall x\in S_1$  we have 
$\langle \phi(x)\rangle =\langle \phi(x_3)\rangle=\langle \phi(x_4)\rangle$, in particular 
$\langle \phi(x_2)\rangle =\langle \phi(x_3)\rangle=\langle \phi(x_4)\rangle$.
On the other hand, $\forall x\in S_2$  we have $\langle \phi(x)\rangle =\langle \phi(x_3)\rangle=\langle \phi(x_4)\rangle$,  
which implies that $\langle \phi(x_1)\rangle =\langle \phi(x_3)\rangle=\langle \phi(x_4)\rangle$.
Combining the two  observations it is straightforward to infer that  $\langle \phi(x_1)\rangle = \langle \phi(x_2)\rangle$.  
In the case when two points are separated by $d(x_1,x_2) \geq 2 R_{\text{H}}$, auxiliary 
intermediate points shall to be introduced and a sequence of inferences of the kind 
presented above has be performed. This completes the proof. 

Therefore, when all degrees of freedom are placed in the same ground state the 
corresponding configuration of space is expected to be ideally homogeneous. This is, 
of course, under the assumption that the angular momentum variables associated 
with the rotational invariance play a significant role in description of the quantum 
state of the gravitational field. As we discuss in the Appendix, this is the case 
at least in one of the most promising approaches to quantum gravity. 

\section{Initial conditions for inflation} \label{Inflation}

Using the formula (\ref{qdefformulagen}) one can hypothesize that the cosmological evolution 
is associated with transition:
\begin{equation}
q \approx i \ (\text{UV}) \rightarrow q = 1\ (\text{IR}).
\end{equation}
The evolution is associated with \emph{decondensation} in which the value of $j_{\text{max}}$ 
increases from $j_{\text{max}}=0$ to $j_{\text{max}}\rightarrow \infty$. One can say that new quantum 
states are released while evolution of the Universe is progressing, with the decrease of energy density. 

An interesting issue to consider is if the described mechanism can provide proper initial conditions 
at the beginning of inflation, which usually require a huge order of homogeneity. Let us discuss this issue 
in more details. We denote $t_{\text{I}}$ to be the time at which inflation starts and $t_{\text{Pl}}$ 
corresponds to the Planck epoch in which  $q\rightarrow i$. Value of the Hubble factor at the beginning 
of inflation is $H(t_{\text{I}})$ and the associated value of the Hubble radius $R_{\text{H}}(t_{\text{I}})=1/H(t_{\text{I}})$.
For the inflation to start the homogeneity scale $L$ at the beginning of  inflation $L(t_{\text{I}})$ must
satisfy $L(t_{\text{I}}) \geq R_{\text{H}}(t_{\text{I}})$. The homogeneity scale at $t_{\text{I}}$ and $t_{\text{Pl}}$ 
can be related via $L(t_{\text{I}}) = L(t_{\text{Pl}}) \frac{a(t_{\text{I}})}{a(t_{\text{Pl}})}$, which leads 
to the condition
\begin{equation}
L(t_{\text{Pl}})   \geq  \frac{a(t_{\text{Pl}})}{a(t_{\text{I}})} \frac{H(t_{\text{Pl}})}{H(t_{\text{I}})}R_{\text{H}}(t_{\text{Pl}}),
\end{equation}
with $R_{\text{H}}(t_{\text{Pl}})\approx l_{\text{Pl}}$. For the barotropic matter, the above inequality can be written as 
\begin{equation}
L(t_{\text{Pl}})   \geq  \left(  \frac{\rho_{\text{Pl}}}{\rho_{\text{I}}} \right)^{\frac{1+3w}{6(1+w)}}  l_{\text{Pl}},
\end{equation}
where $w$ is the barotropic index. The problem of initial homogeneity is associated with the fact that 
for pre-inflationary period (where $1+3w>0$) we and  $\rho_{\text{Pl}} \gg {\rho_{\text{I}}}$ we have 
$L(t_{\text{Pl}})  \gg   l_{\text{Pl}}$. The homogeneity scale at the Planck epoch has to be much greater than 
the Planck length (or equivalently the Hubble radius). Thanks to the condensation mechanism introduced 
in Sec. \ref{Condensation} such condition has, however, chance to be satisfied. Even if one do not support 
the criticism of the Ref. \cite{Ijjas:2013vea}, the initial homogeneity of the Universe can be still considered as a 
support of the naturalness of the Inflationary paradigm. We would also like to emphasize that our approach 
is independent of the particular model of inflation.

Furthermore, some preliminary considerations regarding primordial perturbations can be made. 
Let us namely notice that due to the ``Planckian pixels'', the number of degrees of freedom associated 
with the Hubble sphere is roughly
\begin{equation}
N \approx \frac{A_{\text{H}}}{l_{\text{Pl}}^2} = 4\pi \frac{R^2_{\text{H}}}{l^2_{\text{Pl}}}.
\end{equation}
Assuming the equilibrium configuration, the average energy is 
\begin{equation}
\langle E \rangle = N \frac{1}{2}T = \frac{A_{\text{H}}T}{2l_{\text{Pl}}^2}. 
\end{equation}
This allows us to quantify thermal fluctuations of the energy:
\begin{equation}
\sigma^2_E = \langle E \rangle^2 - \langle E^2 \rangle = T^2 \frac{\partial \langle E \rangle}{\partial T} =  
\frac{A_{\text{H}}T^2}{2l_{\text{Pl}}^2}.
\end{equation}
In consequence, the relative fluctuations of the energy of the Hubble sphere are 
\begin{equation}
\delta_E := \frac{\sigma_E}{\langle E \rangle} \sim \frac{1}{\sqrt{N}} \sim \frac{H}{m_{\text{Pl}}}. 
\end{equation}
Considering the energy density $\rho  = \frac{E}{V_{\text{H}}}$ in a fixed Hubble volume $V_{\text{H}}$ we obtain
\begin{equation}
\delta_{\rho} := \frac{\sigma_\rho}{\langle \rho \rangle}  \sim \frac{H}{m_{\text{Pl}}},
\end{equation}   
and the power of perturbations 
\begin{equation}
|\delta_{\rho}|^2  \sim \left(\frac{H}{m_{\text{Pl}}}\right)^2. 
\end{equation}   
The predicted amplitude of the fluctuations is, therefore, expected to be in qualitative 
agreement with the inflationary power spectrum.  However, the result is very preliminary 
and further more sophisticated investigations are required to confirm if the correct 
inflationary powers spectrum can be recovered.  In particular, as shown in Ref. \cite{Magueijo:2006fu}, 
the considerations similar to the one presented above may lead to nearly scale invariant 
spectrum of primordial perturbations. 

\section{Thermodynamics of the Hubble horizon} \label{Thermodynamics}

\begin{figure}[ht!]
\centering
\includegraphics[width=12cm,angle=0]{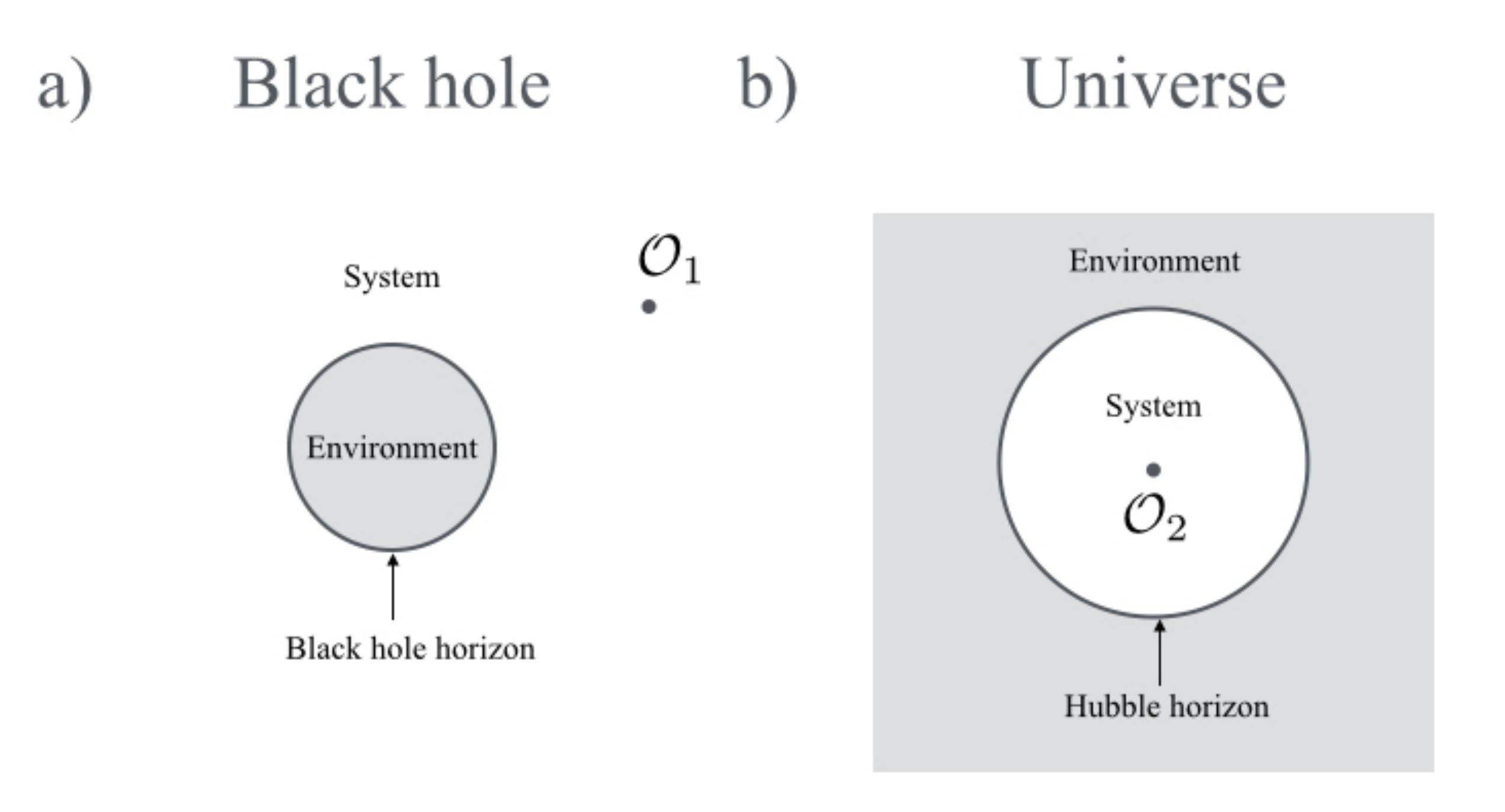}
\caption{a) For an observer $\mathcal{O}_1$, the black hole entropy is a measure of lack of knowledge
about the environmental degrees of freedom, hidden under the black hole horizon. b) In cosmology, for 
any observer $\mathcal{O}_2$ there is always a Hubble horizon, which defines boundary between the 
system (interior of the Hubble sphere) and the environment (exterior of the Hubble sphere).} 
\label{horizons}
\end{figure}

As we have discussed, the $q$-deformations can be interpreted as a consequence of the finite 
angular resolution associated with the Planck scale ``pixels'' at the Hubble sphere. The quantum 
nature of the Hubble sphere indicates that there is a finite entropy associated with the area of the 
Hubble sphere (see also \cite{Davies:1988dk,Davis:2003ye,Mathew:2013vsa}). The entropy is a 
measure of observer's lack of information about the state of the \emph{environment}. From the definition, the 
environmental degrees of freedom are those which are inaccessible to the observer. In the cosmological 
context, the interior of the Hubble sphere can be called a \emph{system} and the region outside of the 
Hubble radius is the environment. The situations is quite the opposite to the case of black holes, 
where the Bekenstein-Hawking entropy 
\cite{Bekenstein:1973ur,Hawking:1974sw}
\begin{equation}
S_{\text{BH}} = \frac{1}{4} \frac{A}{l^2_{\text{Pl}}}
\end{equation}
is associated with lack of access to the information stored under the black hole horizon for an 
observer located outside of the black hole (see Fig. \ref{horizons}). 

Following the holographic principle \cite{tHooft:1993dmi}, let us initially assume that the entropy associated 
with the Hubble horizon is the same as in case of the black hole. At the microscopic level the entropy 
may be explained by associating elementary degrees of freedom with every  Planck scale ``pixel'' in 
the spirit of the \emph{if from bit} conjecture \cite{Wheeler}. The number of such elementary 
Planck cells is $N \approx A/l^2_{\text{Pl}}$. Assuming that state of a Planck cell is encoded by a 
single bit, the number of microstates corresponding to the configuration with a fixed area is given by 
$\Omega = 2^{N} $. Based on this, the Boltzmann entropy of the horizon is 
$S = \ln \Omega = N \ln 2 \sim A/l^2_{\text{Pl}}$.
 
Let us now consider the system to be defined as the interior of the Hubble sphere of the radius 
$R_{\text{H}}$ and the environment to be the exterior (see Fig. \ref{horizons}). The corresponding 
entropy associated with the horizon can be then expressed as 
\begin{equation}
S_{\text{H}} = sS_{\text{BH}} =  \frac{s}{4} \frac{4\pi R_{\text{H}}^2}{l^2_{\text{Pl}}}
= \frac{s}{4} \frac{4\pi}{l^2_{\text{Pl}} H^2} \label{HorizonEntropy}
\end{equation}
where we introduced $s=\left\{-1,0,1 \right\} $. The choice the factor $s$ depend on direction of 
the entropy transfer between the system and environment. In the case of the black holes (for an
observer located outside of the horizon) the increase of the horizon area is associated with the 
increase of the entropy of the environment (volume outside of the horizon is decreasing). The 
entropy is transferred from the region of decreasing volume to the region of increasing volume.  

In the cosmological context, it is worth considering the ratio between the Hubble radius $R_{\text{H}}$ 
and some physical scale $L \propto a$:
\begin{equation}
\frac{R_{\text{H}}}{L} \sim \frac{1}{a|H|} \sim |H|^{-\frac{1+3w}{3(1+w)}}. 
\end{equation} 
In the expanding Universe, for  $w>-\frac{1}{3}$, the horizon size is increasing with respect to the physical 
scales. We, therefore, expect that the entropy of the Hubble volume is increasing and in consequence $s=1$. 
Simply, more degrees of freedom enter the system so $dS_{\text{H}} > 0$. In turn, for $w<-\frac{1}{3}$ the 
Hubble radius in shrinking with respect to physical scales and the entropy of the system is expected to 
decrease, therefore $s=-1$.  Here, the degrees of freedom leave the system leading to $dS_{\text{H}}<0$. 
For $w=-\frac{1}{3}$ the  $\frac{R_{\text{H}}}{L} =$ const and in consequence we have to fix $s=0$ since 
there is no entropy transfer in this case.  Furthermore, the volume enclosed by the Hubble sphere is 
$V_{\text{H}}= \frac{4}{3}\pi R^3_{\text{H}}$, which allow to write energy in this volume as 
$U=V_{\text{H}} \rho$, where $\rho$ is the energy density. 

The system under considerations (interior of the Hubble sphere) satisfies the first law of thermodynamics
\begin{equation}
dU=TdS_{\text{H}}-pdV_{\text{H}}, \label{firstlaw}
\end{equation}
and the second law of thermodynamics
\begin{equation}
dS_{\text{H}}+dS_{\text{env}} \geq 0,  \label{seconlaw}
\end{equation}
where $dS_{\text{env}}$ is the entropy change of the environment. The weak inequality reduces 
to equality if irreversible processes are not present in the system. In particular, if the entropy change is only by 
the exchange of heat between the system and environment, then $dQ_{\text{H}}=-dQ_{\text{env}}$
and in consequence $dS_{\text{H}}+dS_{\text{env}}=0$. In such a case the entropy of the system 
($S_{\text{H}}$) can be reduced by the cost of increase of the entropy of the environment ($S_{\text{env}}$). 
Such behavior is one of the characteristics of the \emph{open systems}, which allows to departure 
from the state of thermal equilibrium \cite{Prigogine}.  In what follows, we will focus on
the case of the entropy exchange by the heat transfer, such that the entropy (\ref{HorizonEntropy}) 
can be used for both the system and the environment (but with opposite signs).

Furthermore, the two laws of thermodynamics are accompanied by the equation of state (EOS),
which in the considered case is played by the Friedmann equation: 
\begin{equation}
H^2=\frac{8 \pi l^2_{\text{Pl}}}{3}\rho,  \label{Friedmann}
\end{equation}
where we neglected the curvature term. Using the expression for the area of the Hubble horizon 
$A = 4\pi /H^2$, the Friedmann EOS can be written as 
\begin{equation}
\frac{3}{2l^2_{\text{Pl}}}=A \rho.  \label{EOS}
\end{equation}
Applying this EOS to the first law of thermodynamics (\ref{firstlaw}), together with 
$U=V_{\text{H}} \rho= \frac{1}{4l_{\text{Pl}}^2} \sqrt{\frac{A}{\pi}}$, expession for the 
derivative of $S_{\text{H}}$ can be obtained: 
\begin{equation}
dS_{\text{H}} = \frac{\sqrt{A}}{4 \sqrt{\pi} T}\left( \frac{1}{2 l^2_{\text{Pl}} A } + p\right) dA,
\end{equation}
which shows that $S_{\text{H}}$ can be written as a function of a single variable $A$.  
It remains to express $T$ and $p$ in terms of $A$. For this purpose, let us rewrite the 
first law of thermodynamics (\ref{firstlaw}) into the form 
\begin{equation}
\dot{\rho}-3 \frac{\dot{H}}{H}(\rho+p) = \frac{T}{V_{\text{H}}} \frac{dS_{\text{H}}}{dt}. \label{dSdt}
\end{equation}
Note that this equation is significantly different from the case in which one consider thermodynamics of a 
local, unspecified region of the Universe. In our case, the Eq. \ref{dSdt} is \emph{not} equivalent to the 
continuity equation. On the other hand, the energy density $\rho$ satisfies the \emph{local} conservation 
laws, which lead to the continuity equation 
\begin{equation}
\dot{\rho}+3H(\rho+p) = 0. \label{cont}
\end{equation}
Eqs. \ref{dSdt}, \ref{cont} may be re-written as:
\begin{equation}
3H(\rho+p)\frac{\rho+3p}{2\rho} =  \frac{T}{V_{\text{H}}} \frac{dS_{\text{H}}}{dt} \, . \label{HubbCont}
\end{equation}
In agreement with our previous analysis, the equation (\ref{HubbCont}) predicts that $dS_{\text{H}}=$ const 
for $p=-\frac{1}{3}\rho$ but also for $p=-\rho$ due to the fact that the entropy is a function of $A$, 
which remains constant in the de Sitter universe. 

With the use of the Friedmann equation (\ref{Friedmann}) and employing (\ref{HorizonEntropy}) we can 
now solve the equation (\ref{HubbCont}) such that the expression for the pressure can be found:
\begin{equation}
p = -\frac{1}{3}\rho+\frac{sTH}{2l_{\text{Pl}}^2}.
\end{equation}
This equation implies that in the expanding Universe with $w > -\frac{1}{3}$ ($s=1$) the pressure 
$p > -\frac{1}{3}\rho$, as expected.  On the other hand, for  $w <  -\frac{1}{3}$ ($s=-1$) we have 
$p < -\frac{1}{3}\rho$. The cosmic acceleration can be, therefore, associated with the horizon entropy
decrease. More specifically, accelerated expansion and the decrease of entropy are tautology for the 
Hubble horizon entropy given by the Bekenstein-Hawking formula with negative sign. 

Taking the barotropic equation of state $p=w\rho$ the expression for the 
temperature of the thermal bath (environment) can be found
\begin{equation}
T =\frac{(3w+1)}{2 s} \frac{H}{2\pi}.  \label{HawkingTemp}
\end{equation}
One can notice that for de Sitter case ($w=-1$ and $s=-1$) the know expression for 
the de Sitter horizon temperature $T =\frac{H}{2\pi}$ \cite{Pollock:1989pn} is correctly recovered.

The above considerations concerned the case of barotropic fluid. But to be more realistic,   
let us now study the $\Lambda$CDM cosmology and try to reconstruct the expression for 
the entropy function. In this case, the total energy density is a sum of contributions from 
(pressureless) dark matter and cosmological constant $\Lambda$:
\begin{equation}
\rho = \rho_{DM}+\rho_{\Lambda} = \frac{\rho_{DM,0}}{a^3} + \frac{\Lambda}{8\pi G}. \label{energydensity}
\end{equation}  
Contribution of radiation has been neglected. However, such fraction is expected in particular due to the 
cosmological analogue of the Hawking radiation. While such contribution is expected to be rather marginal, 
situations when the cosmological Hawking radiation may play significant role cannot be ruled 
out (see Ref. \cite{Barrau:2014kza}). 

Applying (\ref{energydensity}) the formula (\ref{dSdt}) we can find that 
\begin{equation}
\frac{dS_{\text{H}}}{dA} = \frac{1}{16 \pi l^2_{\text{Pl}}} \frac{H}{T} \left( 1 -\frac{\Lambda}{H^2} \right). \label{dSdA}
\end{equation}
In case of de Sitter universe with $T=\frac{H}{2\pi}$ and $H^2 = \frac{\Lambda}{3}$, the Eq. (\ref{dSdA}) gives 
\begin{equation}
\frac{dS}{dA} = - \frac{1}{4l^2_{\text{Pl}}}, 
\end{equation}
which correctly leads to the expression 
\begin{equation}
S_{\text{H}} =S_0 - \frac{A}{4 l^2_{\text{Pl}}}. \label{SDeSitter}
\end{equation}
Note that the minus sign in this equation corresponds to $s=-1$ in the Eq. (\ref{HorizonEntropy}). 
In the general case, we do not know what relation between $T$ and $H$ is. However, because of 
dimensional reason, we expect that the linear relation is preserved such that $T= c H$, with some 
dimensionless constant $c$. With the use of this, the Eq.  (\ref{dSdA}) can be written as 
\begin{equation}
\frac{dS_{\text{H}}}{dA} = \frac{1}{16 \pi l^2_{\text{Pl}} c} \left(1 -\frac{\Lambda}{4\pi} A \right),  \label{dSHdA}
\end{equation}
which can be solved to 
\begin{equation}
S_{\text{H}} = S_0 +\frac{A}{4l^2_{\text{Pl}}} \frac{\left(1 -\frac{\Lambda}{8\pi} A \right)}{4\pi c}. \label{LCDMEntropy}
\end{equation}
In the $\Lambda\rightarrow 0$ limit the expression correctly reduces to dust case ($w=0$) for which
$S_{\text{H}}$ is given by Eq. \ref{HorizonEntropy} with $s=1$ and $c=\frac{1}{4\pi}$ (from Eq.  \ref{HawkingTemp}). 
De Sitter limit is a little more tricky since in this case the entropy is a constant  because the area of the 
horizon is equal $A = \frac{12\pi}{\Lambda}$.  This can be taken into account by choosing the 
integration constant $S_0$ in Eq. \ref{LCDMEntropy} to be 
\begin{equation}
S_0 = \frac{9\pi}{4 l^2_{\text{Pl}}\Lambda}. \label{S0}
\end{equation}
This guarantees that in de Sitter limit the entropy (\ref{LCDMEntropy}) correctly reduces to the 
Bekenstain-Hawking entropy of de Sitters space, with negative sign (as expected for $w=-1$), i.e. 
\begin{equation}
S_{dS} = - \frac{A}{4 l^2_{\text{Pl}}} = - \frac{3\pi}{l^2_{\text{Pl}}\Lambda}.
\end{equation}

There is, however, a problem with the $S_0$ given by Eq. \ref{S0} since this factor 
diverges in the $\Lambda\rightarrow 0$ limit. But, one has to keep in mind that what 
physically matters is not the absolute value of entropy but the entropy change, 
which is a subject of measurements. The entropy change is always well defined
and the issue with the limit $\Lambda\rightarrow 0$ does not spoil behavior of the 
entropy change expected in this limit. This is because derivative of $S_{\text{H}}$ 
can be always taken before the $\Lambda\rightarrow 0$ limit.   

The Eq. \ref{LCDMEntropy} indicates that the $\Lambda$CDM model can be seen 
a result of thermodynamics of the Hubble volume with entropy given by the linearly 
corrected Bekenstein-Hawking entropy.  Worth stressing is that the finiteness of the 
entropy is a consequence of the Planck scale discreteness of the Hubble horizon. 
Furthermore, corrections to the Bekenstein-Hawking entropy arise in various approaches 
to quantum gravity, such as Loop Quantum Gravity \cite{Meissner:2004ju}. However, 
the corrections are typically of the logarithmic type and agreement of the entropy 
(\ref{LCDMEntropy}) with some models of the quantum gravitational degrees of freedom 
is to be examined.  

Moreover, based on Eq. \ref{dSHdA}, one can conclude that the entropy of the system 
(Hubble volume) is decreasing for the $\Lambda$CDM Universe if 
\begin{equation}
\Lambda > \frac{4\pi}{A} = H^2. \label{EntropyCondition}
\end{equation}
Based on the most up to date astronomical observations \cite{Ade:2015xua} we have
\begin{equation}
\Omega_{\Lambda} := \frac{\Lambda}{3H_0^2} \approx 0.69 \pm 0.01, 
\end{equation}
where $H_0$ is the current value of the Hubble factor, which gives 
\begin{equation}
\Lambda \approx  2.07 H_0^2 > H_0^2.
\end{equation}
The condition (\ref{EntropyCondition}) is, therefore, satisfied in the observed Universe,
allowing for the entropy decrease. The presented results suggest that there is relation 
between the cosmic accelerated expansion and the entropy reduction (complexity growth)
in the observable Universe. This possibility will be investigated in more details elsewhere. 

\section{Summary} \label{Summary}

In this article, we have performed analysis of possible cosmologically relevant consequences
of the Planck scale discreteness of the Hubble horizon. Following the results presented in  
Ref. \cite{Bianchi:2011uq}, we have associated the quantum nature of the Hubble horizon with 
the deformations of the rotation symmetry. Mathematically,  this relationships leads to the so-called 
$q$-deformations of the $SO(3)$ or $SU(2)$ groups. Using the fact that the $q$-deformation 
leads to the constraint on the number of irreducible representations, we have shown that in the limit
of Planckian energy densities only the ground states can be occupied. This provides a mechanism 
of generation of primordial isotropy. Then, combining this with the Copernican Cosmological Principle, 
we argued that, homogeneity spanned over many Hubble volumes at the Planck epoch can 
be obtained. This gives a possible resolution of problem of initial homogeneity at the onset of inflation. 

The discreteness of the Hubble horizon leads to a finite entropy function associated with 
heat exchange across the horizon.  In particular, the entropy function may take the form of 
the Bekenstein-Hawking formula. In our studies, we performed thermodynamical analysis 
of the system defined as the interior of the Hubble volume and exterior playing the role of 
environment. We have shown that the entropy transfer can take place in both directions
between the system and the environment. The Hubble volume can be, therefore, perceived as an 
\emph{open system}. In the expanding Universe, for the barotropic index $w>-\frac{1}{3}$
the system gains entropy from the environment. On the other hand, for $w<-\frac{1}{3}$
the system is reducing its entropy. This later case may have profound consequences 
for the increase of complexity in the observed Universe (confront with Ref. \cite{TDComplexity}). 
Furthermore, there is no entropy (heat) transfer for $w=-\frac{1}{3}$. Finally, we have reconstructed 
the form of entropy function for the $\Lambda$CDM model, obtaining Bekenstein-Hawking formula  
with linear correction in the area of the horizon. Such corrections may possibly arise due to 
more detailed counting of quantum states associated with the cosmological horizon. 
The results open new ways to explore relation between the Planck scale effects and 
observationally relevant features of our Universe. 

\section*{Acknowledgements}

MA was supported by the Iuventus Plus grant No. 0290/IP3/2016/74 from the Polish Ministry of 
Science and Higher Education. JM is supported by the Grant DEC-2014/13/D/ST2/01895 of the 
National Centre of Science and by the Mobilno\'s\'c Plus Grant 1641/MON/V/2017/0 of the Polish 
Ministry of Science and Higher Education. 

\section{Appendix: $q$-deformation in Loop Quantum Gravity} \label{LQG}

One of the most studied approaches to the Planck scale physics is the background independent 
Loop Quantum Gravity (LQG). The starting point for LQG is the formalism of  Ashtekar variables for 
which the $A$ and $E$ are canonical fields, satisfying the $\mathfrak{su}(2)$ algebra are considered 
\cite{Ashtekar:1986yd}. The phase space of such the classical GR written in the framework of Ashtekar 
variables is affine. However, while passing to LQG, the connection $A$ is a subject of exponentiation, 
forming a holonomy which is an element of the compact group ${\rm SU}(2)$ \cite{Rovelli:1989za}. 
The fluxes constructed with the use of $E$ are elements of the $\mathfrak{su}(2)$ algebra. 

In the covariant formulation, the LQG is related the so-called Ponzano-Regge model of quantum 
gravity \cite{Barrett:2008wh} which relies on the  ${\rm SU}(2)$ group. As it has been shown for the 
2+1-dimensional Ponzano-Regg models an unbounded value of $j$ leads to the IR divergences 
called \emph{spikes}. At the beginning of 90's of last century, it has been shown first at the level of
purely mathematical considerations (Turaev-Vito model) and then in the work of S.~Mizoguchi and 
T.~Tada \cite{Mizoguchi:1991hk}, that the divergences can be cured if the  $q$-deformation of the 
${\rm SU}(2)$ group is introduced. It was concluded that,  if the deformation with $q \neq 1$ is present, 
the values of $j$ are bounded from above, removing the IR divergences of the theory. Moreover, 
in the $(2+1)$-dim case it has been shown that the deformation parameter $q$ introduces non-vanishing 
cosmological constant into the theory. Namely, one can find that the value of the cosmological constant 
$\Lambda$ is related to the parameter $q$ via the formula 
\begin{equation}
q = e^{i \Lambda l_{\text{Pl}}^2}, \label{qdefformulaMT}
\end{equation}
so that for $\Lambda \rightarrow 0$ the undeformed case is recovered. It is unknown, however, if this 
relation holds in the $(3+1)$-dim case  \cite{Rovelli:2015fwa}. Taking the current value of the Hubble 
radius $\sim 1/\sqrt{\Lambda} \sim 10^{26}$m one obtains to $q \simeq 1 + i 10^{-122}$ and $j_{\text{max}} \sim 10^{122}$. 

Worth stressing is that the formula (\ref{qdefformulaMT}) is consistent with the Eq. (\ref{qdefformulagen})
derived in this article. Namely, the energy density of the cosmological constant is 
\begin{equation}
\rho_{\Lambda} = \frac{\Lambda}{8 \pi l_{\text{Pl}}^2},
\end{equation}
which, when applied to Eq. (\ref{qdefformulagen}), reproduces Eq. (\ref{qdefformulaMT}).

\end{document}